\documentclass[twoside]{ilcws07}
\usepackage[latin1]{inputenc}
\usepackage[dvips]{graphicx,epsfig,color}
\usepackage{wrapfig,rotating}
\usepackage{amssymb,amsmath,array}

\newcommand{\cm}[1]{\ensuremath{ {{\rm cm}^{#1}}}}
\newcommand{\ntp}[2]{\ensuremath{#1\times10^{#2} } }
\newcommand{\asb}[2]{\ensuremath{#1_{\rm #2} }}

\begin{document}

\begin{titlepage}

\thispagestyle{empty}
\def\thefootnote{\fnsymbol{footnote}}       % symbols for footnotes

\begin{center}
\mbox{ }

\end{center}
%\begin{flushright}
%\vspace* {-2.0cm}
%\Large
%\mbox{\hspace{10.2cm} hep-ph/yymmnnn} \\
%\end{flushright}

\vskip 1cm
\hspace*{-2cm}
\begin{picture}(0.001,0.001)(0,0)
\put(,0){
\begin{minipage}{1.2\textwidth}
\begin{center}
\vskip 1.0cm

{\Huge\bf
Simulations of the Temperature
}
\vspace{3mm}

{\Huge\bf
Dependence of the Charge Transfer 
}
\vspace{3mm}

{\Huge\bf
Inefficiency in a High-Speed CCD
}
\vskip 1.5cm
{\LARGE\bf 
Andr\'e Sopczak

\bigskip

\Large Lancaster University, UK}

\vskip 1.5cm
%\centerline{\Large \bf Abstract}
{\Large \bf Abstract}

\end{center}
\end{minipage}
}
\end{picture}

\vskip 8cm
\hspace*{-1cm}
\begin{picture}(0.001,0.001)(0,0)
\put(,0){
\begin{minipage}{\textwidth}
\Large
\renewcommand{\baselinestretch} {1.2}
Results of detailed simulations of the charge transfer inefficiency
of a prototype serial readout CCD chip are reported. The effect of radiation damage on the chip
operating in a particle detector at high frequency at a future accelerator is studied, specifically the creation of
two electron trap levels, 0.17\,eV and 0.44\,eV below
the bottom of the conduction band. Good agreement is found between
simulations using the ISE-TCAD DESSIS program and an analytical
model for the former level but not for the latter. Optimum operation is predicted to be
at about 250\,K where the effects of the traps is minimal; this being
approximately independent of readout frequency in the range 7--50\,MHz. The work has been
carried out within the Linear Collider Flavour Identification
(LCFI) collaboration  in the context of the International Linear
Collider (ILC) project.
\renewcommand{\baselinestretch} {1.}

\normalsize
\vspace{2.5cm}
\begin{center}
{\sl \large
Presented at LCWS'07,
Linear Collider Workshop 2007 and the International Linear Collider meeting 2007,
DESY, Hamburg, Germany, 2007, \\
to be published in the proceedings.
\vspace{-6cm}
}
\end{center}
\end{minipage}
}
\end{picture}
\vfill

\end{titlepage}

\clearpage
\thispagestyle{empty}
\mbox{ }
\newpage
\setcounter{page}{1}
\pagestyle{plain}

\title{
Simulations of the Temperature Dependence 
of the Charge Transfer Inefficiency in a 
High-Speed CCD}

\author{Andr\'e Sopczak\thanks{Email: andre.sopczak@cern.ch}
, on behalf of the LCFI collaboration
% Optional short acknowledgment: remove next line if non-needed
% DO NOT MODIFY THE FOLLOWING '\vspace' ARGUMENT
\vspace{.3cm}\\
% Addresses and institutions (remove "1- " in case of a single institution)
Lancaster University
}
%%***********************************************************************
% END OF AUTHORS INFORMATION AREA
%***********************************************************************

\maketitle

\begin{abstract}
Results of detailed simulations of the charge transfer inefficiency
of a prototype serial readout CCD chip are reported. The effect of radiation damage on the chip
operating in a particle detector at high frequency at a future accelerator is studied, specifically the creation of
two electron trap levels, 0.17\,eV and 0.44\,eV below
the bottom of the conduction band. Good agreement is found between
simulations using the ISE-TCAD DESSIS program and an analytical
model for the former level but not for the latter. Optimum operation is predicted to be
at about 250\,K where the effects of the traps is minimal; this being
approximately independent of readout frequency in the range 7--50\,MHz. The work has been
carried out within the Linear Collider Flavour Identification
(LCFI) collaboration  in the context of the International Linear
Collider (ILC) project.
\end{abstract}

\section{Introduction}
Particle physicists worldwide are working on the design of a high
energy collider of electrons and positrons (the International Linear
Collider or ILC) which could be operational sometime around 2019. Any
experiment exploiting the ILC will require a high performance vertex
detector to detect and measure short-lived particles, yet be
tolerant to radiation damage for its anticipated lifetime. One candidate
is a set of concentric cylinders of
Charge-Coupled Devices (CCDs), read out at a frequency of around 50\,MHz.

It is known that CCDs suffer from both surface and bulk radiation damage. However,
when considering charge transfer losses in buried channel devices
only bulk traps are important. These defects create energy levels
between the conduction and valence band, hence electrons may be
captured by these new levels. These electrons are also emitted back
to the conduction band after a certain time.

It is usual to define a Charge Transfer Inefficiency
(CTI), which is the fractional loss of charge after transfer across one pixel.
An initial charge $Q_0$ after being transported across
$m$ pixels is reduced to
$Q_m=Q_0(1-{\rm CTI})^m$.
For CCD devices containing many pixels, CTI values 
around 10$^{-5}$ are not negligible.

The CTI value depends on many parameters, some related to the trap
characteristics such as: trap energy level, capture cross-section,
and trap concentration (density). Operating conditions also affect
the CTI as there is a strong temperature dependence on the trap
emission rate and also a variation of the CTI with the readout
frequency. Other factors are also relevant, for example the mean
occupancy ratio of pixels (1\% for a 50\,MHz readout is assumed here), which influences the fraction of filled
traps in the CCD transport region.

Previous studies have been
reported in~\cite{Janesick,Stefanov,Ursache,Brau,Sopczak}.
The novel features of this work are detailed 2D simulations using real device
geometry without approximations for the charge storage volume and transport.
\begin{figure}[htp]
\vspace*{-5.5mm}
\begin{minipage}{0.49\textwidth}
\includegraphics[width=0.7\columnwidth,clip]{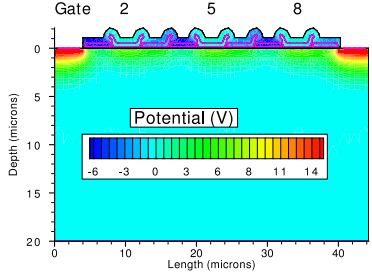}
\end{minipage}\hfill
\begin{minipage}{0.49\textwidth}
\caption{\label{fig:trap} \footnotesize Detector structure and the potential under the gates
after initialization. The signal charge is injected under gate 2 and is moved to the right.
There are three gates for each pixel.}
\end{minipage}
\vspace*{-3mm}
\end{figure}

\section{Simulations}
\vspace*{-2mm}

The UK Linear Collider Flavour Identification (LCFI)
collaboration~\cite{LCFI,Greenshaw} has been studying a serial readout device
produced by e2V Technologies, with a manufacturer's designation
`CCD58'. It is a 2.1\,Mpixel, three-phase buried-channel CCD with
12\,$\mu$m square pixels.

Simulations of a simplified model of this device have been performed with
the ISE-TCAD package (version 7.5), particularly the DESSIS program
(Device Simulation for Smart Integrated Systems). It contains an
input gate and an output
gate\footnote{These separate the area under study from the
input drain and output diffusion pn junctions.}, a substrate contact and nine further gates
(numbered 1 to 9) which form the pixels. Each pixel consists of 3
gates but only one pixel is important for this study---gates 5, 6
and 7. 
The simulation is essentially two dimensional and assumes a 1\,$\mu$m 
device thickness (width) for calculating densities.
Thus the model is equivalent to a short, thin slice of one column of CCD58 with rectangular pixels
12\,$\mu$m long by 1\,$\mu$m wide. The overall length and depth are 44\,$\mu$m and 20\,$\mu$m 
respectively (Fig.~\ref{fig:trap}).

\begin{figure}[htp]
\vspace*{-0.3cm}
\begin{minipage}{0.69\textwidth}
\includegraphics[width=0.49\columnwidth,height=3.7cm,clip]{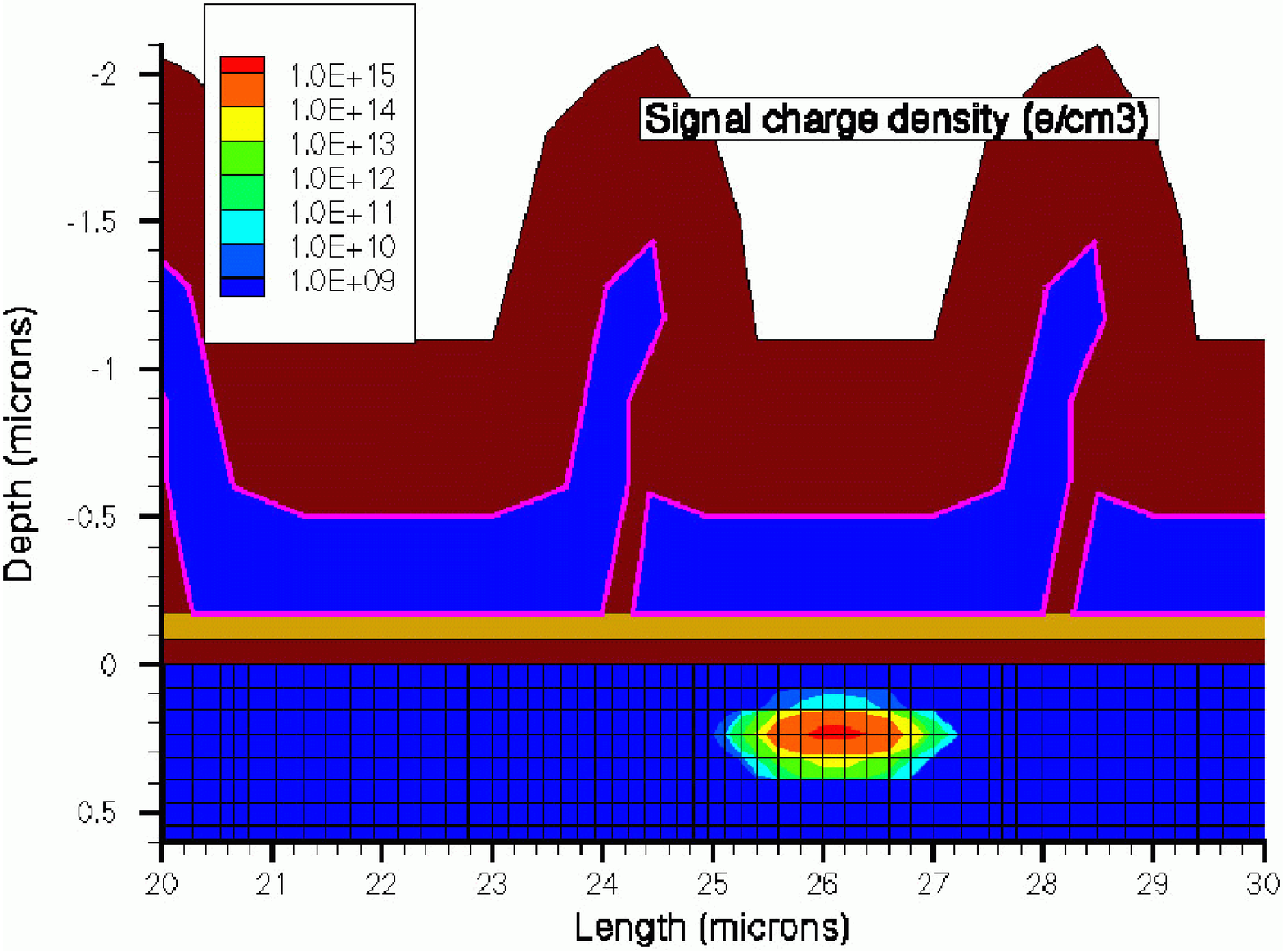}
\includegraphics[width=0.49\columnwidth,height=3.7cm,clip]{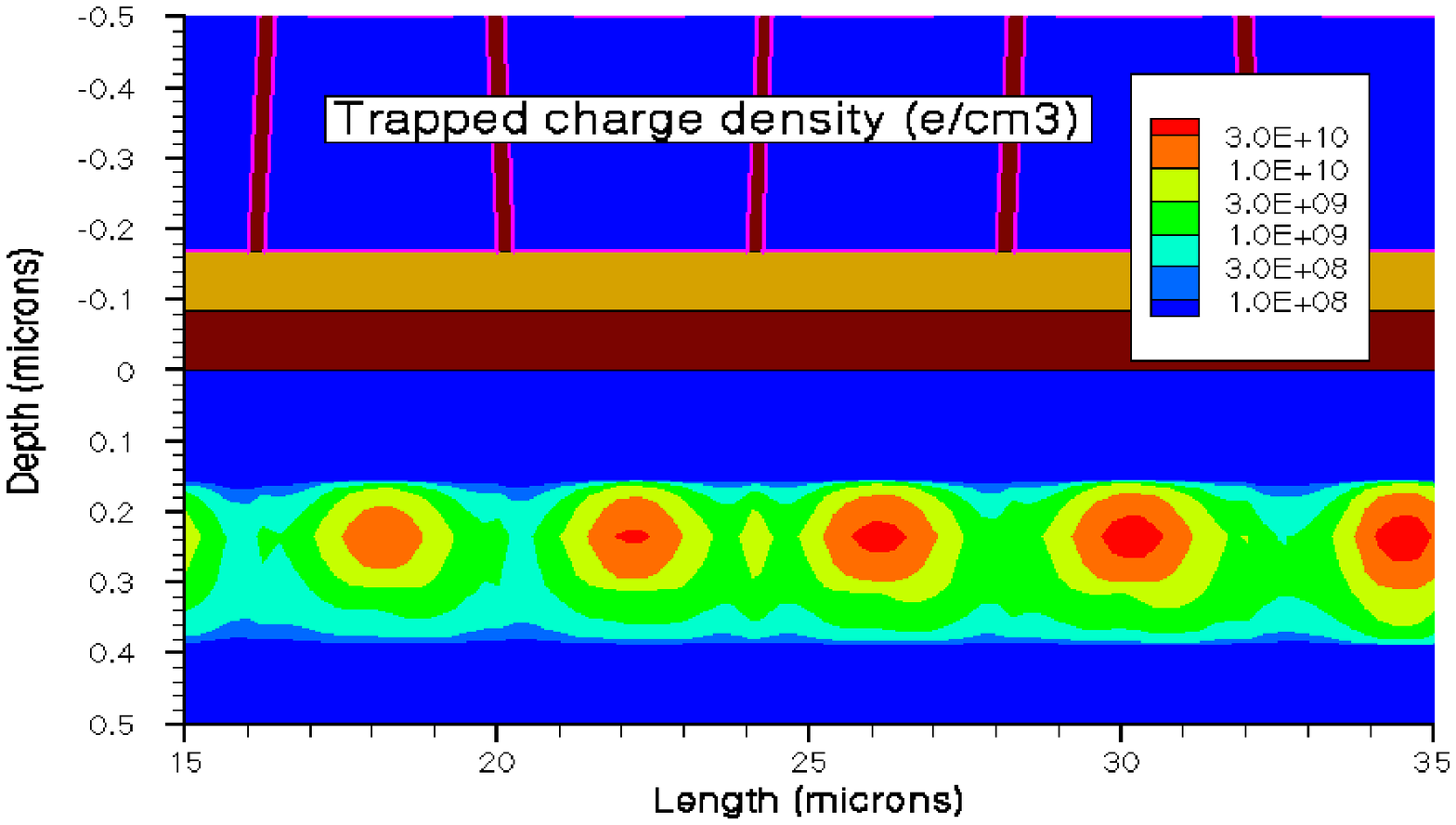}
\end{minipage}
\begin{minipage}{0.29\textwidth}
\caption{\label{fig:transport} \footnotesize 
Left: signal charge density. 
The mesh size varies between 0.1 and 0.3 microns. 
During the analysis an integration under each gate is performed. 
Right: trapped charge density from transfer of signal charge
at a time when the signal packet has passed under all 
}
\vspace*{0.5mm}
\end{minipage}
\footnotesize 
the gates. The trapped charge density \mbox{decreases} from the right to the left due to emission. 
The time the charge spends under the gates is much longer than the 
time spent in the gaps (which is of the order of a nanosecond), 
therefore the trapped charge density is much higher under the gates
in comparison with the region between the gates.
The legend box refers to the region with positive depth values. 
At negative depth values are an oxide layer, a nitride layer, 
polysilicon gates and finally an oxide layer. No metal is shown.
\end{figure}

Parameters of interest are the readout frequency, up to 50\,MHz, and
the operating temperature between 120\,K and 300\,K although
simulations have been done up to 500\,K. The charge in transfer and
the trapped charge are shown in Fig.~\ref{fig:transport}.

The signal charge used in the simulation is chosen to be similar to
the charge generated by a minimum ionising particle (MIP), amounting
to about 1620 electron-hole pairs\footnote{This number has to be
divided by 12 because the charge is assumed to be distributed over
the whole pixel but the model has only 1/12th of the true pixel
volume.} for CCD58. DESSIS has a directive for generating heavy ions
and this is exploited to create the charges. The heavy ion is made
to travel in a downwards direction starting at 1.2\,$\mu$m below
gate 2 at 1\,$\mu$s before charge transfer begins. This provides
ample time for the electrons to be drawn upwards to the transport
channel which is 0.25\,$\mu$m beneath the gate electrodes.

\subsection{Calculating CTI}
\vspace*{-1mm}

Charge Transfer Inefficiency is a measure of the fractional loss of
charge from a signal packet as it is transferred over a pixel, or
three gates. After DESSIS has simulated the transfer process, a 2D
integration of the trapped charge density distribution is performed
independently to give a total charge under each gate.
The CTI for transfer over one gate is equivalent to
$CTI=\frac{e_{T}-e_{B}}{e_{S}}, \label{eqn:gatecti}$
where
$e_{S}$ = number of electrons in the signal packet,
$e_{B}$ = number of background trapped electrons prior to signal packet transfer,
$e_{T}$ = number of trapped electrons under the gate, after signal transfer across gate.
 In this way the CTI is normalised for each gate.
The determinations of the trapped charge take place for gate $n$
when the charge packet just arrives at gate $n+1$. If the
determination were made only when the packet has cleared all three
gates of the pixel, trapped charge may have leaked out of the traps.\footnote{Since some of this leaked charge might rejoin the signal packet now under the next gate, this procedure may slightly overestimate the CTI.}

The total CTI (per pixel) is determined from gates 5, 6 and 7, hence
$CTI=\sum_{n=5}^7 \frac{e_{T}-e_{B}}{e_{S}} \label{eqn:pixelcti},$
where $n$ is the gate number.  The background charge is taken as the
trapped charge under gate 1 because this gate is unaffected by the
signal transport when the charge has just passed gates 5, 6 or 7.

\vspace*{-1mm}
\subsection{0.17\,eV and 0.44\,eV traps} 
\vspace*{-1mm}

This CTI study, at nominal clock voltage, focuses only on
the bulk traps with energies 0.17\,eV and 0.44\,eV below the bottom
of the conduction band. These will be referred to simply as the
0.17\,eV and 0.44\,eV traps. An incident particle with sufficient
energy is able to displace an atom from its lattice point leading
eventually to a stable defect. These defects manifest themselves as
energy levels between the conduction and valence band, in this case
the energy levels 0.17\,eV and 0.44\,eV; hence electrons 
may be captured by these levels. The 0.17\,eV trap is an
oxygen vacancy defect, referred to as an A-centre defect.
The 0.44\,eV trap is a phosphorus-vacancy defect---an E-centre
defect---that is, a result of the silicon being doped with
phosphorus and a vacancy manifesting from the displacement of a
silicon atom bonded with the phosphorus atom \cite{Stefanov}.

In order to determine the trap densities for use in simulations, a
literature search on possible ILC radiation backgrounds and trap
induction rates in silicon was undertaken. The main expected
background arises from e$^+$e$^-$ pairs with an average energy of
10\,MeV and from neutrons (knocked out of nuclei by synchrotron
radiation).

Table~\ref{tab1} shows results of background simulations of e$^+$e$^-$ pairs
generation for three proposed vertex detector designs (from three
ILC detector concepts).

\begin{table}[htb]
\vspace*{-3mm}
\begin{minipage}{0.44\textwidth}
\begin{tabular}{|c | c| c| c|}
\hline Simulator & SiD & LDC & GLD \\
\hline CAIN/Jupiter & 2.9 & 3.5 & 0.5 \\
GuineaPig & 2.3 & 3.0 & 2.0 \\
\hline
\end{tabular}
\end{minipage}\hfill
\begin{minipage}{0.54\textwidth}
\caption{\footnotesize\label{tab1} Simulated background results for three
different detector scenarios. The values are hits per square
centimetre per e$^+$e$^-$ bunch crossing. SiD is the Silicon
Detector Concept~\cite{SiD}, LDC is the Large Detector
Concept~\cite{LDC} and GLD is the Global Linear collider
Detector~\cite{GLD}. }
\end{minipage}
\vspace*{-2mm}
\end{table}

Choosing the scenario with the highest expected background, that is
the LDC concept, where the innermost layer of the vertex detector
would be located 14\,mm from the interaction point, one can estimate
an e$^+$e$^-$ flux around 3.5\,hits/cm$^2$/bunch crossing which
gives a fluence of 0.5$\times 10^{12}$\,e/cm$^2$/year. In the case
of neutrons, from two independent studies, the fluence was estimated
to be 10$^{10}$\,n/cm$^2$/year~\cite{Maruyama} and 1.6$\times
10^{10}$\,n/cm$^2$/year~\cite{Vogel}.

Based on the literature~\cite{Marconi, robbins, robbins-roy,
walker,wertheim,suezawa, saks, srour,fretwurst}, the trap densities
introduced by 1\,MeV neutrons and 10\,MeV electrons have been
estimated with two established assumptions: the electron trap
density is a linear function of dose, and the dose is a linear
function of fluence. A summary is given in Table~\ref{tab2}.

\begin{table}
\begin{minipage}{0.64\textwidth}
\begin{tabular}{| l | l| l |}\hline
 Particle type & 0.17\,eV (\cm{-3}) & 0.44\,eV (\cm{-3})  \\ \hline
 10 MeV e$^-$&   \ntp{3.0}{11} & \ntp{3.0}{10}\\
 \hphantom{0}1 MeV n &   \ntp{(4.5\ldots 7.1)}{8} & \ntp{(0.7\ldots 1.1)}{10}
 \\ \hline
 total    &   \ntp{3.0}{11} & \ntp{4.1}{10}\\ \hline
\end{tabular}
\end{minipage}\hfill
\begin{minipage}{0.34\textwidth}
\caption{\label{tab2}\footnotesize Estimated densities of
traps after irradiation for one year. For neutrons, the literature
provides two values.}
\end{minipage}
\end{table}
\noindent The actual trap concentrations and electron capture
cross-sections used in the simulations are shown in Table~\ref{tab3}.

\begin{table}
\begin{minipage}{0.59\textwidth}
\begin{tabular}{| c | l| l| l |}\hline
$\asb{E}{t}-\asb{E}{c}$ (eV)
 & Type & $C$ (\cm{-3}) &$\sigma$ (\cm{2})\\ \hline
0.17 & Acceptor & \ntp{1}{11} & \ntp{1}{-14}\\
0.44 & Acceptor & \ntp{1}{11} & \ntp{3}{-15}\\ \hline
\end{tabular}
\end{minipage}\hfill
\begin{minipage}{0.39\textwidth}
\caption{\label{tab3}\footnotesize Trap concentrations
(densities) and electron capture cross-sections as used in the
DESSIS simulations.}
\end{minipage}
\end{table}

\subsection{Partially Filled Traps}
Each electron trap in the semiconductor material can either be {\em
empty\/} (holding no electron) or {\em full\/} (holding one
electron). In order to simulate the normal operating conditions of
CCD58, partial trap filling was employed in the simulation (which
means that some traps are full and some are empty) because the
device will transfer many charge packets during continuous
operation.

In order to reflect this, even though only the transfer of a single
charge packet was simulated, the following procedure was followed in
all cases. During an initial 98\,$\mu$s period, the gates ramp
up and all the traps are filled. The gates are biased in such a way so that charge moves to the output
drain. The device is then in a fully normal biased state and corresponds to the situation of a charge packet having just passed through the pixel under investigation. Since another charge packet does not arrive immediately, a 2\,$\mu$s waiting
time\footnote{This waiting time corresponds to the mean time between the arrival of charge packets from a 1\% mean pixel
occupancy with a 50\,MHz readout frequency and to larger values for lower frequencies.} is introduced before readout clocking is started. During this period some of the traps become empty. The
test charge is generated 1\,$\mu$s after the start of this waiting period so that
1\,$\mu$s, later when the waiting ends, there is a signal packet sitting under gate 2 just at the time when the three sinusoidally varying voltages (clock phases) are applied to cause the
transfer of the produced signal charge packet through the device.

\section{Simulation Results}

The CTI dependence on temperature and readout frequency was explored.

\subsection{0.17\,eV traps}

\begin{figure}[htp]
\vspace*{-3mm}
\begin{minipage}{0.74\textwidth}
\includegraphics[width=0.49\columnwidth,height=3.5cm,clip]{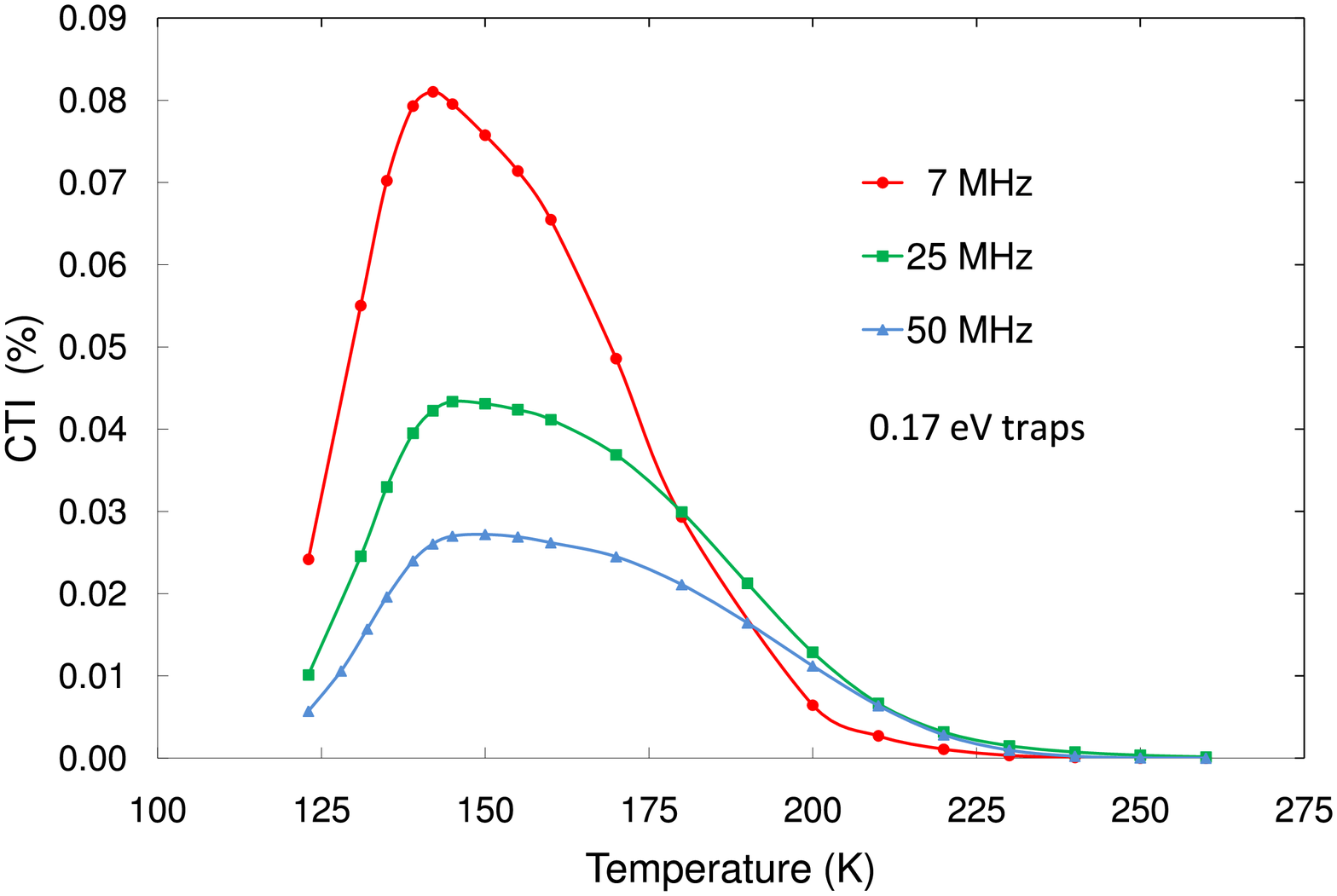}
\includegraphics[width=0.49\columnwidth,height=3.5cm,clip]{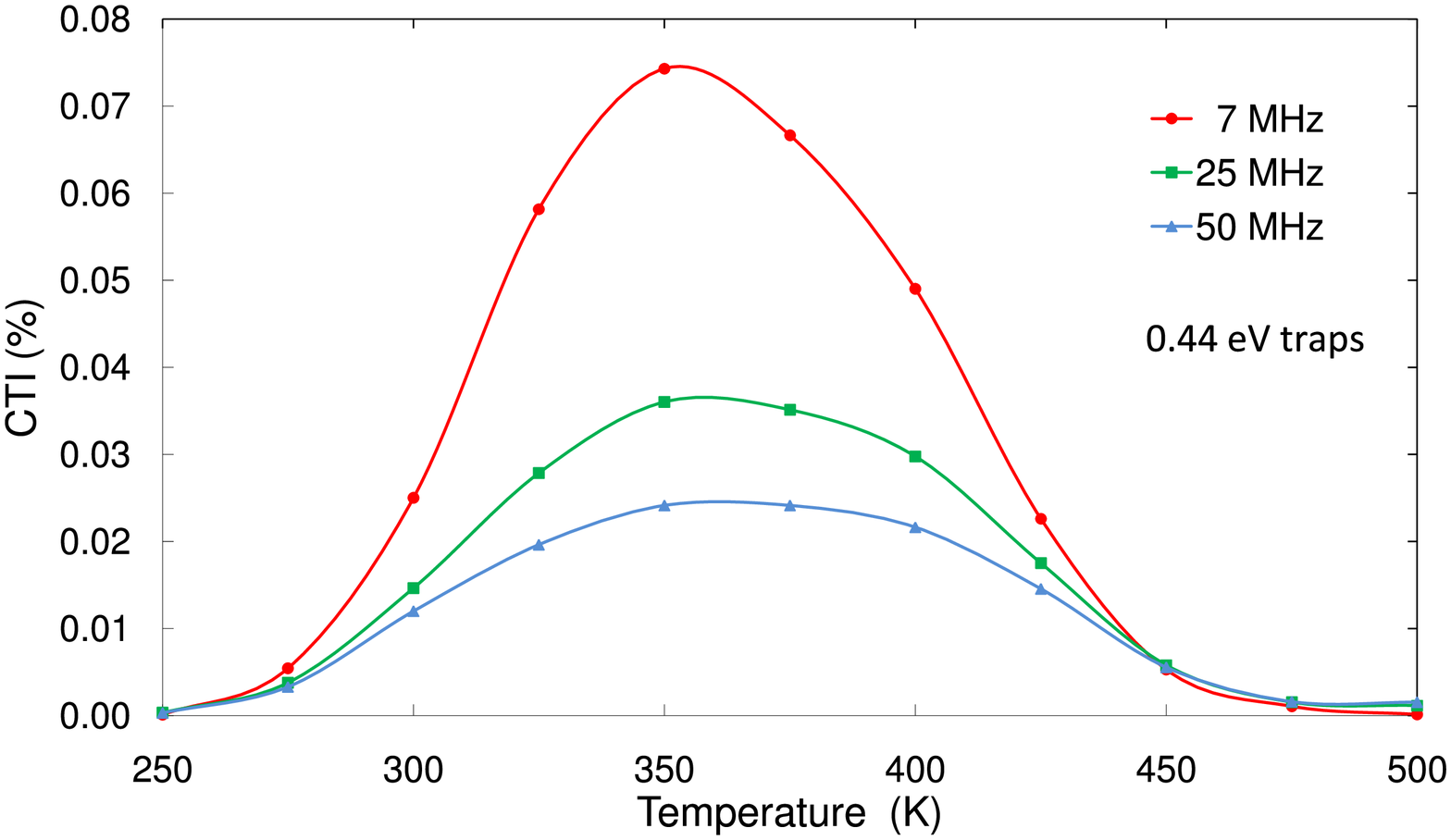}
\end{minipage}\hfill
\begin{minipage}{0.24\textwidth}
\caption{\footnotesize \label{fig:allfreq_0_17eV}CTI values versus temperature for simulations with
0.17\,eV (left) and 0.44\,eV (right) partially filled traps at clocking frequencies 7, 25 and
50\,MHz.}
\end{minipage}
\vspace*{-1mm}
\end{figure}

Figure~\ref{fig:allfreq_0_17eV} shows the CTI for simulations with
partially filled 0.17\,eV traps at different frequencies for
temperatures between 123\,K and 260\,K, with a nominal clock voltage
of 7\,V.

A peak structure can be seen. For 50\,MHz, the peak is at 150\,K
with a CTI of $27 \times 10^{-5}$. The peak CTI is in the region
between 145\,K and 150\,K for a 25\,MHz clock frequency and with a
value of about $43 \times10^{-5}$. This is about 1.6 times bigger
than the charge transfer inefficiency at 50\,MHz. The peak CTI for
7\,MHz occurs at about 142\,K, with a maximum value of about $81
\times 10^{-5}$, an increase from the peak CTI at 50\,MHz $(27
\times 10^{-5})$ by a factor of about 3 and an increase from the
peak CTI at 25\,MHz $(43 \times 10^{-5})$ by a factor of nearly 2.
Thus CTI increases as frequency decreases. For higher readout
frequency there is less time to trap the charge, thus the CTI is
reduced. At high temperatures the emission time is so short that
trapped charges rejoin the passing signal.

\subsection{0.44\,eV traps}

Simulations were also carried out with partially filled 0.44\,eV
traps at temperatures ranging from 250\,K to 500\,K. This is because
previous studies~\cite{Sopczak} on 0.44\,eV traps have shown that
these traps cause only a negligible CTI at temperatures lower than
250\,K due to the long emission time and thus traps remain fully
filled at lower temperatures. The results are also depicted in
Fig.~\ref{fig:allfreq_0_17eV}.
The peak CTI is higher for lower frequencies with little temperature
dependence of the peak position.

\subsection{0.17\,eV and 0.44\,eV traps together}

\begin{figure}[htp]
\begin{minipage}{0.49\textwidth}
\includegraphics*[width=0.9\columnwidth,height=4cm,,clip]{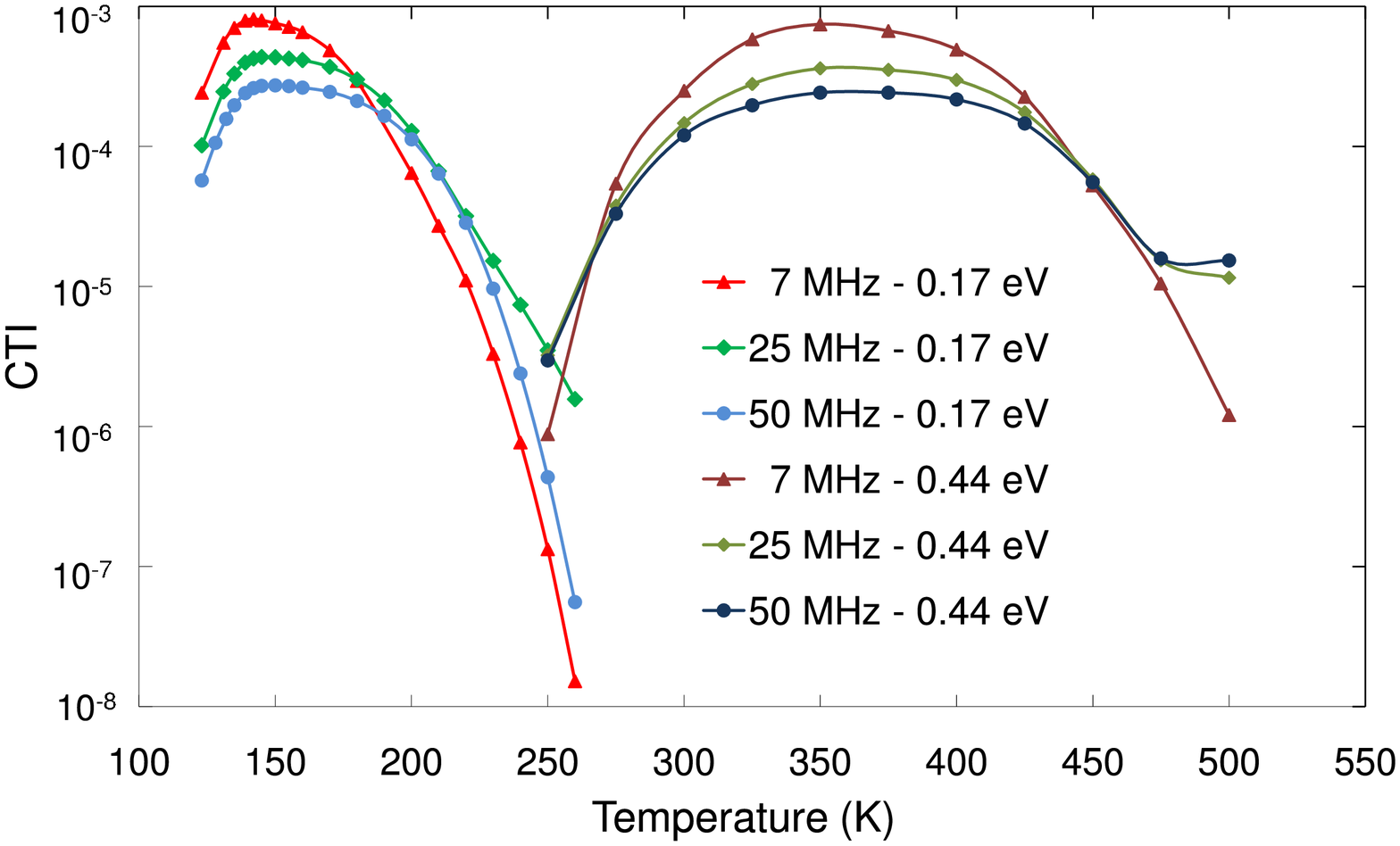}
\end{minipage}\hfill
\begin{minipage}{0.49\textwidth}
\caption{\footnotesize\label{fig:all_log}CTI values versus temperature for simulations for partially
filled 0.17\,eV and 0.44\,eV traps at frequencies
7, 25 and 50\,MHz (logarithmic
scale).}
\end{minipage}
\vspace*{-5mm}
\end{figure}

The logarithmic scale plot (Fig.~\ref{fig:all_log}) of the
simulation results at the different frequencies and trap energies
clearly identifies an optimal operating temperature of about 250\,K.

\section{Comparisons with an Analytical Model}
The motivation for introducing an analytical
model is to understand the underlying physics through making
comparisons with the DESSIS simulations. This might then allow predictions of CTI for other CCD geometries without requiring a full simulation.

\subsection{Capture and emission time constants}
The charge transfer inefficiency is modelled by a differential equation
in terms of the different time constants and temperature dependence
of the electron capture and emission processes. In the electron
capture process, electrons are captured from the signal packet and
each captured electron fills a trap. This occurs at a rate determined by a capture time constant
$\tau_{\rm c}$. The electron emission process is described by the
emission of captured electrons from filled traps back to the
conduction band, and into a second signal packet at the emission
rate determined by an emission time constant $\tau_{\rm e}$.

Following the treatment by Kim~\cite{Kim}, based on earlier work by Shockley, Read and Hall~\cite{ShockleyReadHall},
a defect at an energy $E_{\rm t}$ below the bottom of the conduction
band, $E_{\rm c}$, has time constants
\vspace*{-1mm}
\begin{equation}
\tau_{\rm c} = \frac{1}{\sigma_{\rm e} \nu_{\rm th} n_{\rm s}} ~~~~~~
\tau_{\rm e} = \frac{1}{\sigma_{\rm e} \chi_{\rm e} \nu_{\rm th}
N_{\rm c}}\exp{\left(\frac{E_{\rm c} - E_{\rm t}}{k_{\rm
B}T}\right)} 
\end{equation}
where
$\sigma_{\rm e}$ = electron capture cross-section,
$\chi_{\rm e}$ = entropy change factor by electron emission,
$\nu_{\rm th}$ = electron thermal velocity,
$N_{\rm c}$ = density of states in the conduction band,
$k_{\rm B}$ = Boltzmann's constant,
$T$ = absolute temperature,
and $n_{\rm s}$ = density of signal charge packet.
It is assumed that $\chi_{\rm e}=1$.

At low temperatures, the emission time constant $\tau_{\rm e}$ can
be very large and of the order of seconds. The charge shift time for one gate,
$t_{\rm sh}=1/(3{\rm f})$, where
${\rm f}$ is the readout frequency,
is of the order of nanoseconds. A larger $\tau_{\rm e}$ means that a
trap remains filled for much longer than the charge shift time.
Further trapping of signal electrons is not possible and,
consequently, CTI is small at low temperatures. A peak occurs
between low and high temperatures because the CTI is also small at
high temperatures. This manifests itself because, at high
temperatures, the emission time constant decreases to become
comparable to the charge shift time so trapped electrons rejoin
their signal packet.

\subsection{Charge Transfer Model}
The model by Hardy et
al.~\cite{hardy} considers the effect of a single trapping level and includes only the emission time constant in the
following differential
equation
${\mathrm{d}n_{\rm t}}/{\mathrm{d}t}=-{n_{\rm t}}/{\tau_{\rm e}}$
where $n_{\rm t}$ is the density of filled traps.
The traps are
initially filled for this model and $\tau_{\rm c}\ll t_{\rm sh}$.

When  $\tau_{\rm c}\gg t_{\rm sh}$ and to be consistent with the DESSIS simulation (that uses partially filled traps), this model can be adapted by the
 use of the capture time constant. The solution of this differential equation leads to an
estimator of the CTI:

\vspace*{-2.5mm}
\begin{eqnarray}
CTI=\left(1-e^{-t_{\rm sh}/\tau_{\rm c}}\right)\frac{3N_{\rm
t}}{n_{\rm s}}\left(e^{-t_{\rm join}/\tau_{\rm e}}-e^{-t_{\rm
emit}/\tau_{\rm e}}\right) \label{eqn:cti5}
\end{eqnarray}
where
$N_{\rm t}$ is the density of traps,
$t_{\rm emit}$ is the total emission time from the previous packet, the mean
waiting time between charge packets related to the mean occupancy of pixels in the device,
and $t_{\rm join}$ is the time period during which the charges can join the parent charge packet.
This definition is for the CTI for a single trap level. The factor of
three appears since there is a sum over the three gates that make up
a pixel. (The Hardy model solution does not have the terms inside the leftmost bracket.)

\begin{figure}[htp]
\begin{minipage}{0.49\textwidth}
\includegraphics[width=\columnwidth,clip]{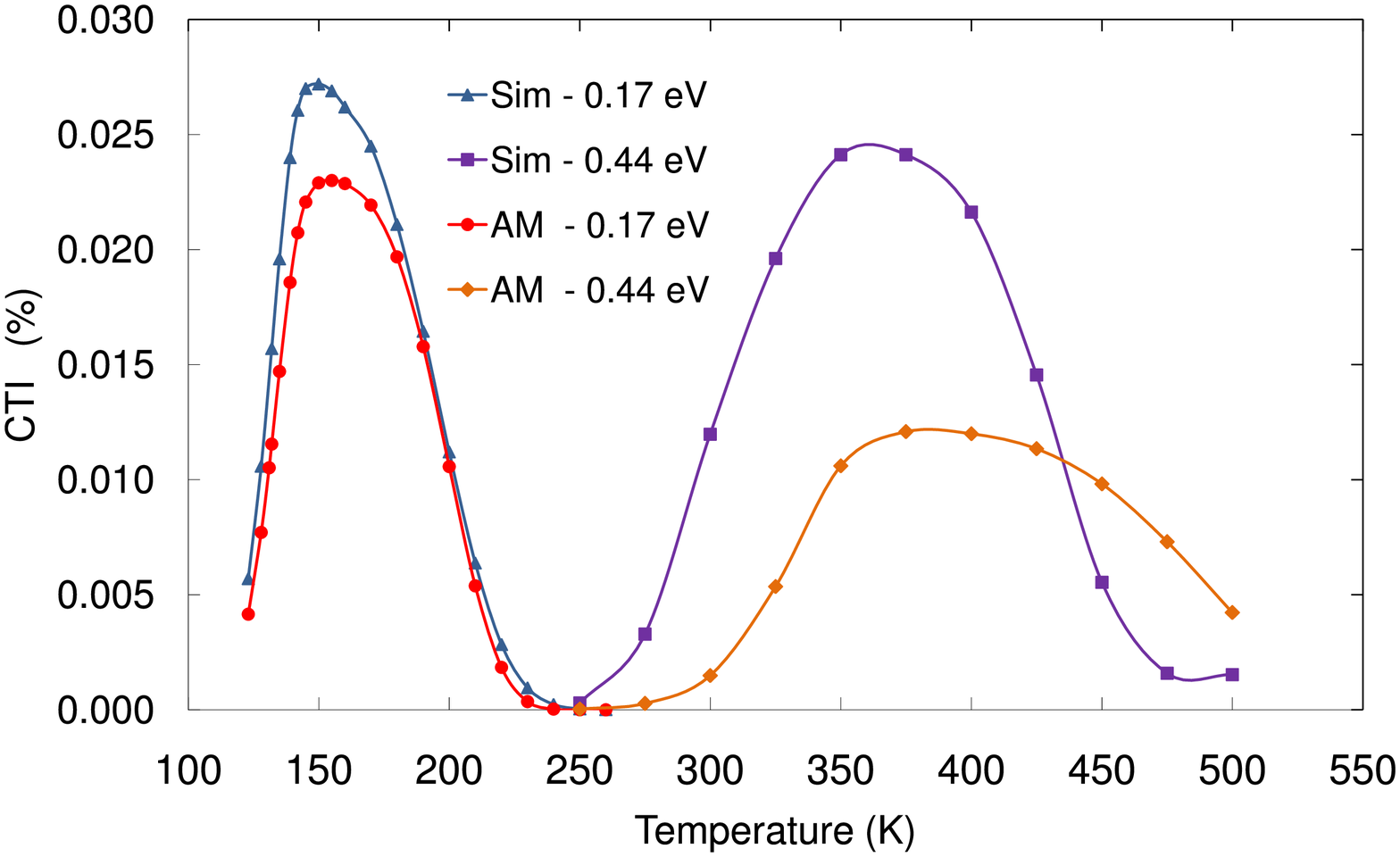}
\end{minipage}\hfill
\begin{minipage}{0.49\textwidth}
\caption{\footnotesize \label{fig:ISE_v_IM}CTI values versus temperature for
simulations for 0.17\,eV and 0.44\,eV partially filled traps at
clocking frequency 50\,MHz. Comparison of the Analytical Model (labelled AM) with the
full DESSIS simulation (Sim).}
\end{minipage}
\end{figure}

Figure~\ref{fig:ISE_v_IM} compares the full DESSIS simulation for
0.17\,eV and 0.44\,eV traps and clocking frequency of 50\,MHz to this
Analytical Model. It emphasises the good agreement between the model
and full simulations at temperatures lower than 250\,K with 0.17\,eV
traps, but shows a disagreement at higher temperatures for the
0.44\,eV traps.

However it
is clear that there are limitations with the Analytical Model. They
could relate to a breakdown of the assumptions at high temperatures,
to ignoring the precise form of the clock voltage waveform, or to
ignoring the pixel edge effects. Further studies are required.

\section{Conclusions and Outlook}

The Charge Transfer Inefficiency (CTI) of a CCD device has been
studied with a full simulation (ISE-TCAD DESSIS) and compared with an
analytical model.

Partially filled traps from the 0.17\,eV and 0.44\,eV trap levels
have been implemented in the full simulation and variations of the
CTI with respect to temperature and frequency have been analysed.
The results confirm the dependence of CTI with the readout
frequency. At low temperatures ($<250$\,K) the 0.17\,eV traps
dominate the CTI, whereas the 0.44\,eV traps dominate at higher
temperatures.

Good agreement between simulations and an Analytical Model
has been found for 0.17\,eV traps but not for 0.44\,eV traps. This
shows the limitations of the model with respect to the full
simulation.

The optimum operating temperature for CCD58 in a high radiation
environment is found to be about 250\,K for clock frequencies in the range 7 to 50\,MHz.
However CCD58 is not really suited to high speed readout and attempts to make laboratory measurements have given inconsistent results. So in order to meet the demanding readout requirements for a vertex detector at the ILC,
interest has now moved to an alternative CCD design with
Column-Parallel (CP) and 2-phase readout. Our prototype CP-CCD has recently operated at 45\,MHz. Thus our involvement with serial readout devices will probably now cease but the experience gained with DESSIS and building analytical models will transfer to our studies of CP-CCDs.

\section*{Acknowledgments}
This work is supported by the Particle Physics and Astronomy
Research Council (PPARC) and Lancaster University. The Lancaster
authors wish to thank Alex Chilingarov, for helpful discussions, and
the particle physics group at Liverpool University, for the use of
its computers.

% IF YOU DO NOT USE BIBTEX, USE THE FOLLOWING SAMPLE SCHEME FOR THE REFERENCES
% ----------------------------------------------------------------------------
\renewcommand{\baselinestretch} {0.98}

% ****************************************************************************
% END OF BIBLIOGRAPHY AREA
% ****************************************************************************

\end{document}